\documentclass{article} % For LaTeX2e
\usepackage{iclr2027_conference,times}

\usepackage{amsmath,amsfonts,bm}

\def\eqref#1{equation~\ref{#1}}
\def\1{\bm{1}}

\DeclareMathAlphabet{\mathsfit}{\encodingdefault}{\sfdefault}{m}{sl}
\SetMathAlphabet{\mathsfit}{bold}{\encodingdefault}{\sfdefault}{bx}{n}

\usepackage{hyperref}
\usepackage{url}
\usepackage{booktabs}
\usepackage{graphicx}
\usepackage{threeparttable}
\usepackage{tabularx}
\usepackage{multirow} 

\title{Think Before You Accept: \\ Can Written Justification Reduce Uncritical Uptake of AI Writing Suggestions?}

\author{%
  Yan Tao$^{1}$, Jennifer Meyer$^{2}$ \& René F. Kizilcec$^{1}$ \\
  $^{1}$Cornell University, Ithaca, USA \\
  $^{2}$University of Vienna, Vienna, Austria 
}

\iclrfinalcopy % Uncomment for camera-ready version, but NOT for submission.
\begin{document}

\maketitle

\begin{abstract}
Generative AI can offer students useful feedback, but its value depends on judging which suggestions are accurate and relevant. Prior research shows that strategic friction during human-AI interactions can promote critical uptake, but how to effectively implement such friction in academic contexts remains unclear. We examine whether requiring students to justify decisions to accept or reject AI suggestions can mitigate uncritical uptake in academic writing. In a randomized experiment embedded in a course activity ($N=129$), students wrote a data analysis proposal, received mixed-quality AI revision suggestions, and decided whether to accept or reject them. Students required to provide written justifications were 24 percentage points less likely to adopt flawed suggestions (65\% vs. 41\%), with no reduction in acceptance of sound suggestions (81\% vs. 86\%). However, thematic analysis revealed superficial engagement in the justification task and gaps in metacognitive monitoring and domain knowledge.
\end{abstract}

\section{Introduction}
Writing is a fundamental competence that is essential for academic and professional success across nearly all disciplines and professions~\citep{graham2019changing, cho2007scaffolded}. Effective academic writing involves an iterative process of goal-oriented evaluation and revision, in which students often need detailed, personalized feedback to guide their improvement~\citep{graham2019changing, flower1981cognitive, graham2015formative}. With the growing availability of large language models (LLMs), college students are increasingly turning to these tools for writing and editing advice~\citep{openai2025ai, black2025university}. Research has shown that, with appropriate prompting, LLMs can provide real-time feedback that supports writing improvement across a range of tasks, from abstracts and short essays to longer academic papers~\citep{meyer2024using, liang2024can, dai2024assessing, glusing2026llm}. Instructor feedback retains distinct strengths in offering more contextually grounded advice~\citep{hyland2026efl, lo2025stretching}, better prioritization of the most important and actionable issues for students to address next~\citep{steiss2024comparing}, and more personal and empathetic guidance~\citep{wang2024chatgpt}. However, producing such feedback is time consuming, limiting how much and how often instructors are able to provide it~\citep{paris2022instructors}. LLMs may therefore complement instructor feedback by providing immediate, iterative support throughout the writing process and identifying issues that instructors may not have the time to address comprehensively~\citep{dai2024assessing}.

A key concern surrounding LLM-generated writing feedback is students’ uncritical acceptance of AI suggestions. Evidence from recent studies suggests that while some students engage with AI suggestions through deliberate evaluation and decision making, others adopt suggestions more mechanically, with limited evaluation of their appropriateness~\citep{du2026exploring, mujeeb2026beyond, alghamdi2026metacognitive}. This tendency is amplified by how contemporary AI writing tools present feedback. Rather than surfacing one issue at a time, commercial assistants typically return a long, itemized list of suggested revisions followed by a single question asking if it should apply all of them (encouraging this behavior by ending with ``just say the word''). This default design creates an asymmetry in interaction cost that favors wholesale acceptance because accepting all suggestions requires little effort, whereas selectively evaluating and adopting individual suggestions requires substantially more. %Evaluating suggestions individually is a significant departure from existing practice. 
Such uncritical uptake of AI suggestions is concerning for academic writing because it can have costs at multiple levels. First, the quality of AI suggestions is not consistently reliable, and blindly accepting them may negatively affect writing quality. Although LLM feedback has been found to be comparable to human teacher feedback for surface-level issues such as grammar and clarity, its quality is contested for content and argumentation~\citep{craven2026llm, hyland2026efl}. For example, LLMs may introduce fabricated frameworks or invalid references when providing research and writing assistance~\citep{davis2026citation}, offer decontextualized suggestions that overlook disciplinary or domain-specific needs~\citep{tseng2026context}, or identify writing problems that do not actually exist~\citep{jia2024assessing}. Second, writing is a self-directed process of thinking and learning, and revision constitutes an important part of deepening understanding of a topic or problem~\citep{bereiter1987psychology,flower1981cognitive}. Students who mechanically agree with LLM-generated revision suggestions may lose their opportunities to reflect on, evaluate, and develop their own ideas. Finally, writing is also a process of self-expression through which students communicate their perspectives. Defaulting to AI suggested revisions may therefore constrain students’ rhetorical agency by giving up control over what to say and how to say it~\citep{tseng2026context}.

To address the problem of uncritical uptake, this study examines an instructional intervention that requires students to provide written justifications for accepting or rejecting LLM generated writing suggestions. 
The intervention builds on broader HCI research showing that preserving friction during human-AI interactions can encourage more critical evaluation of AI suggestions~\citep{buccinca2021trust}. We adapt this design principle to academic writing by introducing a pedagogical form of friction: requiring students to articulate the reasoning behind each adoption decision. Beyond slowing down AI uptake, this requirement is designed to prompt self-explanation and elicit metacognitive monitoring of how AI suggestions may or may not contribute to students' ideas, goals, and writing~\citep{manlove2007software}. Similar practices have begun to emerge in higher education. For example, news media has reported instructors in Australia requiring students to maintain a ''\textit{decision log}'' as a primary assessment artifact documenting their adoption decision for each AI output with 50-75 word justifications ~\citep{elkhodr2026students}. Yet, despite this theoretical grounding and emerging use in practice, little is known about whether requiring written justification can effectively reduce students' uncritical uptake of AI suggestions in academic settings. It also remains unclear how the intervention shapes students' evaluation of AI suggestions and what limitations in their critical evaluation persist. Addressing these questions is important for informing the use of written justification in instructional practice and educational technology design.
%Although requiring students to justify their responses to AI generated content has been incorporated into systematic training for digital critical thinking~\citep{alshehri2026tool}, its effectiveness as a standalone instructional intervention in mitigating uncritical uptake of AI writing suggestions has not been experimentally evaluated using behavioral measures beyond students’ self reported perceptions. 

To examine the effects of the justification intervention, we embedded a randomized experiment within a writing activity in which students received deliberately designed LLM feedback of varying quality. We investigated whether requiring students to justify their decisions changes how they respond to individual AI suggestions, focusing on the following research questions:
\begin{description}
    \item[\textbf{RQ1:}] How does requiring written justification shape students' selective engagement with AI-generated revision suggestions and their discrimination between suggestions of different quality?

    \item[\textbf{RQ2:}] Among students required to justify their decisions, what aspects of flaws in AI-generated revision suggestions do students evaluate, and what do these evaluations reveal about gaps in students' critical AI evaluation?
    %What flaws did students recognize, and what reasoning supported rejection?
    %What did students say when they nevertheless adopted a flawed suggestion?
\end{description}

The contribution of this study is three-fold. First, the study advances understanding of how to mitigate uncritical AI uptake by examining the effects of inserting a written justification step into students' decisions to adopt AI suggestions. Our findings provide causal evidence that requiring justification reduces, though does not eliminate, uncritical uptake of AI suggestions. Our study design isolates the effect of the requirement, not its precise mechanism, but the justifications students wrote indicate that the metacognitive engagement our intervention elicits was often shallow. Second, we provide an example of adapting the HCI principle of strategic friction to educational interventions for effective AI use in learning. Our findings demonstrate the potential of combining interactional friction with pedagogical prompts for metacognitive monitoring, while also revealing gaps that point to opportunities for further interdisciplinary research. Third, the study provides implementation insights for practitioners, including college instructors and AI writing tool designers, seeking to incorporate strategic friction into students' interactions with AI. By examining students' written justifications, we identify persistent gaps in how students evaluate AI suggestions and highlight considerations for designing interventions that better support critical AI use.

\section{Background}
\subsection{The Challenge of Critical Evaluation on AI Writing Suggestions}
While LLMs can provide immediate and detailed feedback on writing across grammar, clarity, organization, content, and argumentation~\citep{liang2024can, dai2024assessing, glusing2026llm}, the quality of such feedback is not always reliable. LLMs may produce feedback that lacks specificity, contains hallucinated information, or fails to account for the broader context and goals of a student's writing~\citep{davis2026citation, tseng2026context, jia2024assessing}. Effective use of AI writing feedback therefore requires students to evaluate individual suggestions and selectively adopt those that are appropriate for their writing goals and task context~\citep{zhan2025students}. However, emerging research suggests that students do not always engage in such critical evaluation. For example, in a case study of Chinese students using LLMs for English academic writing at a private language institute, students reported mechanically clicking “accept, accept” for everything suggested by ChatGPT during revision~\citep{du2026exploring}. Similarly, a study of 156 Pakistani students found that 35\% reported accepting AI suggestions for English academic writing without critical evaluation~\citep{mujeeb2026beyond}.

Several factors may contribute to this tendency toward uncritical uptake. Cognitive biases can lead people to defer to AI outputs. Students may perceive automated writing tools as more knowledgeable or capable than themselves and therefore be more willing to defer to their suggestions~\citep{ranalli2021l2}. This tendency is consistent with automation bias, in which people overrely on automated recommendations and accept them without sufficient independent verification~\citep{goddard2012automation}. The polished and authoritative presentation of LLM generated feedback may further reinforce this tendency by making suggestions appear credible even when they are inaccurate or inappropriate. From a dual-process perspective, people also often favor fast, heuristic-based judgments over slower, more effortful reasoning~\citep{wason1974dual, evans2003two, buccinca2021trust}. The design of current conversational AI tools compounds this asymmetry. When suggestions arrive as a batch followed by a single acceptance question, accepting all of them costs one click, whereas accepting a subset costs an evaluation and an action per item. Fast, uncritical acceptance is thus both cognitively and mechanically cheaper. 
%Together, these cognitive tendencies can make quickly accepting an AI suggestion relatively effortless, whereas determining whether it is accurate, appropriate, and aligned with one's writing goals requires more deliberate evaluation. 

From a learning perspective, metacognitive and self-regulatory processes play an important role in enabling students to resist these cognitive tendencies and engage in deliberate evaluation~\citep{liu2026does, zhao2026self}. Critically evaluating an AI suggestion requires students to monitor their task progress and learning goals, evaluate the suggestion in relation to those goals, and regulate their subsequent actions accordingly. However, students vary in the extent to which they engage in these self-regulatory processes when writing with AI, with research identifying different profiles of metacognitive engagement~\citep{alghamdi2026metacognitive}. Moreover, the ease and immediacy of AI-assisted revision may make it particularly easy for students to repeatedly act on cognitive shortcuts rather than engage in deliberate evaluation. Over time, such repeated deference to AI may reduce students' metacognitive engagement with the writing process, creating a self-reinforcing cycle in which students defer more evaluative decisions to AI, become less attentive to their task situation and learning progress, and have fewer opportunities to train the metacognitive processes needed to critically assess AI suggestions. Reduced metacognitive capacity may, in turn, make the cognitively easier strategy of deferring to AI even more likely~\citep{fan2025beware, noorbehbahani2026ai, risko2016cognitive}.

Together, these factors suggest that uncritical uptake of AI writing suggestions may arise from both students’ cognitive tendency to defer to AI and the self-regulatory and metacognitive demands involved in evaluating AI suggestions. Supporting effective use of AI writing feedback therefore requires interventions that interrupt the cycle of automatically handing evaluative decisions to AI by creating opportunities for students to pause, re-engage their own judgment, and critically evaluate individual suggestions before deciding whether to adopt them. Importantly, the goal should not be to reduce AI reliance indiscriminately, but to promote selective reliance. Research on human-AI decision making suggests that reliance is appropriate when people accept correct AI output and reject incorrect output; thus, the relevant outcome is not simply whether and how often students defer to AI, but how well their reliance tracks the quality of AI suggestions~\citep{raees2026people, schemmer2023appropriate, bansal2021does}.

%Research on human–AI decision making labels this \textit{inappropriate reliance}. Reliance is appropriate when a person accepts correct AI output and rejects incorrect output, so the quantity of interest is not how often someone defers but how well their deference tracks quality~\citep{raees2026people, schemmer2023appropriate, bansal2021does}. Thus, an effective intervention should not simply lower acceptance across the board, which is trading one error for another. We adopt this framing from the HCI literature and treat discrimination between sound and flawed suggestions, rather than overall acceptance, as the outcome that matters for AI-assisted writing.
\subsection{Required Justification as an Intervention for Reducing Uncritical AI Uptake}

Prior research has shown that adding strategic friction to interrupt otherwise fast, heuristic decision making can encourage slower and more deliberate analysis~\citep{lambe2016dual, buccinca2021trust}. In human-AI interactions, \textit{cognitive forcing functions} introducing such friction, such as delaying AI suggestions and requiring users to actively request AI assistance rather than receiving suggestions automatically, have been found to facilitate more critical adoption of AI suggestions~\citep{buccinca2021trust}. However, how such friction can be implemented in academic contexts %, and why some friction works and some does not, 
remains an open question. 

%Vasconcelos and colleagues ~\citep{vasconcelos2023explanations} frame overreliance as a cost-benefit judgment: people engage with AI output when the cost of verifying it is low relative to the effort of solving the problem themselves, and an intervention reduces overreliance when they change that ratio rather than merely add steps. Requiring justification could therefore make accepting a suggestion costly when a student has no reason to accept, reducing \textit{careless} acceptance without reducing acceptance overall.

In the context of academic writing, the stakes of AI reliance extend beyond accuracy. Writers who compose with an opinionated language model shift their expressed views toward the model's~\citep{jakesch2023co}, and writers who accept substantial AI contributions report reduced ownership of the resulting text~\citep{draxler2024ai}. Studies of interactive writing systems note that the point of contact between the writer and the model (when suggestions appear, what the writer must do to take them, etc.) shapes how much of the writing remains the writer's own~\citep{lee2022coauthor}. Our study takes one such point of contact, the moment of accepting or rejecting a suggestion, and investigates what a small change in its cost by adding a cognitive forcing function does to the selectivity and quality of the decision. By introducing an additional evaluative step between receiving an AI suggestion and deciding whether to adopt it, this requirement can potentially interrupt automatic uptake. 
Besides, from a pedagogical perspective, requiring justification may also prompt metacognitive monitoring of students' writing and revision processes. This mechanism resembles prompted self-explanation, an instructional technique that encourages learners to articulate their reasoning and has been shown to support monitoring of understanding, identification of knowledge gaps, and integration of new information with prior knowledge~\citep{bisra2018inducing, chi1989self}. In AI-assisted writing, justifying whether to accept or reject a suggestion may similarly prompt students to monitor their decision making, consider how the suggested change relates to their writing goals and task context, and identify gaps or problems in their reasoning.

Despite these theoretical grounding, empirical evidence on written justification as an intervention for mitigating uncritical AI uptake remains limited. News media has reported higher education practitioners' integration of required justification for AI output adoption decision as a primary assignment requirement~\citep{elkhodr2026students}, but its effectiveness, particularly in academic writing settings, remains unknown. To our knowledge, only Alshehri and colleagues~\citep{alshehri2026tool} have empirically examined adoption decision justification as part of an educational intervention for critical AI use in writing education. Alongside systematic instruction on a guided human-AI collaboration workflow, their intervention required students to maintain AI-use logs documenting the prompts they submitted, outputs they consulted, decisions to accept or reject AI suggestions, and justifications for those decisions. They found improvements in writing fluency and students' self-reported digital critical thinking. However, because written justification was implemented as one component of a broader instructional program, its independent effect cannot be determined. The study also relied on broad learning outcomes and students’ self-reported perceptions rather than their actual responses to individual AI suggestions. Thus, it remains unclear whether and how requiring students to justify their decisions can causally reduce uncritical uptake of AI-generated writing suggestions, particularly whether it helps students discriminate between sound suggestions and those containing AI flaws that should be rejected. Addressing this gap is important for refining instructional practices that aim to promote more critical and self-regulated use of AI writing feedback.

To address this gap, we embedded a randomized experiment within an in-class writing activity in which students received deliberately designed, mixed-quality AI feedback. 
We compared students’ responses across conditions along two behavioral dimensions of critical AI uptake: selectivity of adoption across suggestions of different quality (i.e., whether students adopted all suggestions or selectively adopted some) and quality of adoption decisions (i.e., whether students adopted flawed AI suggestions). This design allowed us to experimentally test whether requiring written justification changes how students evaluate and act on individual AI suggestions, thereby providing behavioral evidence of whether and how this simple, in situ intervention can promote more critical use of AI feedback in post-secondary academic writing.

\section{Method}
\subsection{Study Procedure}
The study was embedded in an in-class activity in an upper-division data science course at a U.S. university, with a focus on applications of data science in education. The course enrolled more than 200 students, including upper-division undergraduate and master’s students, as well as a smaller number of sophomores, from a wide range of academic majors, including information science, computer science, economics, communication, and others. Weekly discussion sections regularly addressed educational topics, including responses to emerging technologies such as generative AI. As part of this broader agenda, the activity provided students with hands-on experience interacting with LLM-generated feedback for academic writing and served as a basis for subsequent in-class peer discussion about how AI feedback should be used in academic writing.

\begin{figure*}[t]
    \centering
    \includegraphics[width=0.9\textwidth]{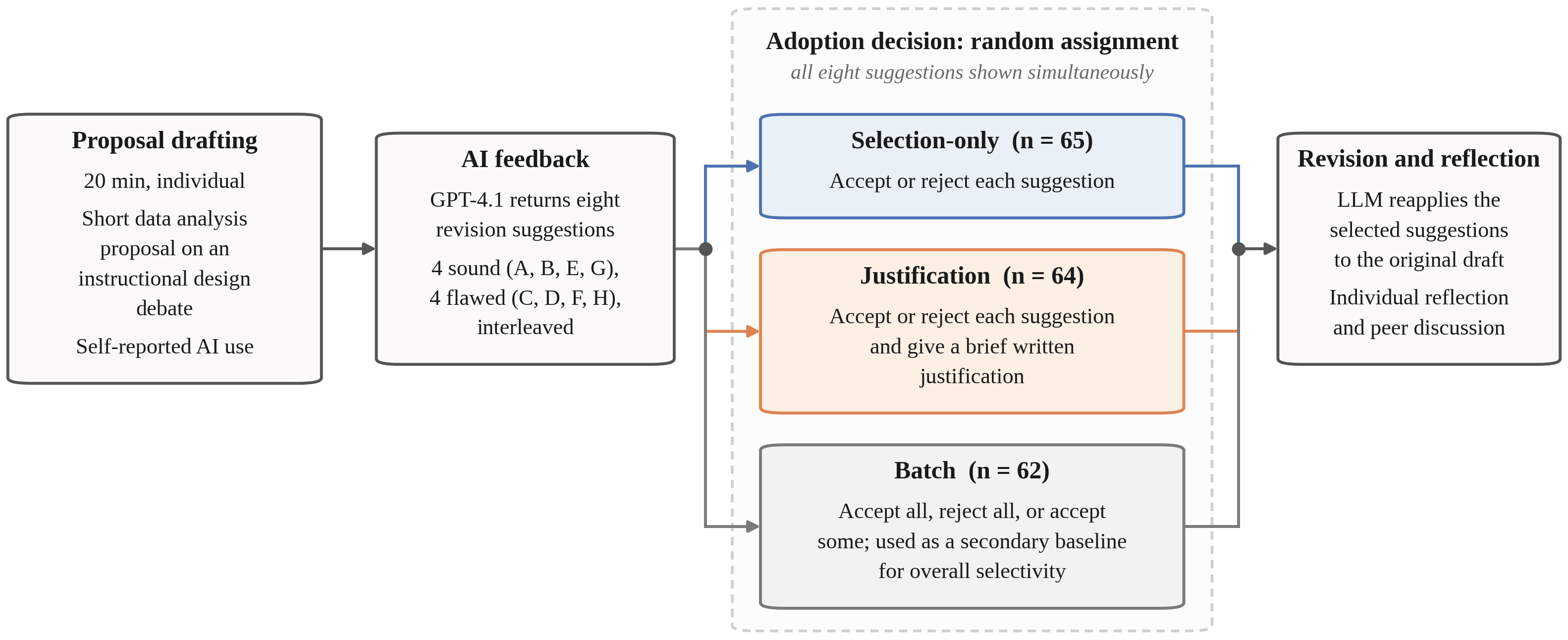}
    \caption{Procedure of the in-class writing activity. Students drafted a short data analysis proposal, received eight LLM-generated revision suggestions of deliberately mixed quality presented simultaneously, and were randomly assigned in approximately equal numbers to one of three conditions ($N = 196$ assigned; five students did not complete the task). The primary comparison is between the selection-only and justification conditions; the batch condition reproduces the accept-all default of commercial AI writing tools and contributes a descriptive estimate of overall selectivity, because a configuration error prevented students who chose ``accept some'' from selecting multiple suggestions. Blue and orange match the condition colors in Figure~\ref{fig:acceptance_rate}.}
    %\Description{A left-to-right flow diagram of four study stages. Stage one, proposal drafting: 20-minute individual task writing a short data analysis proposal on an instructional design debate, with self-reported AI use. Stage two, AI feedback: GPT-4.1 returns eight revision suggestions, four sound and four flawed, interleaved. Stage three, adoption decision, with all eight suggestions shown at once and participants randomly assigned to three conditions: Selection-only (n = 65), accept or reject each suggestion; Justification (n = 64), accept or reject each and give a brief written justification; and Batch (n = 62), accept all, reject all, or accept some. Stage four, revision and reflection: an LLM reapplies the selected suggestions to the original draft, followed by individual reflection and peer discussion.}
    \label{fig:study}
\end{figure*}

The activity procedure is illustrated in Figure~\ref{fig:study}. When introducing the activity, instructors framed it as a hands-on exercise to prepare students for a subsequent class discussion, without explicitly prompting students to critically evaluate AI suggestions. Students first had 20 minutes to write a short data analysis proposal addressing a debate about an instructional design choice. They were asked to explain their position on the debate and propose how educational data could be used to generate empirical evidence to test their position. The proposal was required to include four common components of data analysis proposals in education: (1) a clear, testable hypothesis grounded in learning theory or students’ real world experiences, (2) a clear description of an appropriate data source, (3) an appropriate analytic approach for testing the hypothesis, and (4) a clear structure with minimal grammatical errors. Students were not given specific requirements regarding AI use during this writing stage. After submitting their drafts, they were prompted to disclose whether and how they had used AI during the drafting process.

Students’ drafts were then forwarded to GPT-4.1, which generated eight revision suggestions for each proposal. We selected GPT-4.1 because it was offered by a widely used LLM provider and was among the most cost-effective options available at the time for deployment in a large-enrollment course. The suggestions were deliberately designed to vary in quality to create a structured learning context in which students could practice critically evaluating AI-generated writing feedback. In particular, the flawed suggestions were intentionally constructed to represent common limitations of LLM-generated feedback and to create opportunities for students to identify and discuss problematic suggestions. Similar to using distractors in a multiple-choice question to reveal gaps in students’ knowledge or skills, these flawed suggestions provided concrete cases for identifying gaps in students' critical awareness of AI suggestions. Presenting all students with personalized, mixed-quality suggestions in a fixed sequence also provided a structured basis for follow-up discussion, allowing students to compare their decisions, identify gaps in their evaluation practices, and share effective strategies for critically evaluating AI suggestions. As shown in Table~\ref{tab:feedback_design}, Suggestions C, D, F, and H represented four distinct AI flaw types. Suggestion C provided inaccurate theoretical information, including a fabricated or misapplied theoretical framework accompanied by fictitious references, representing \textit{fabrication}. Suggestion D recommended an irrelevant data source as a way to strengthen the proposed analysis, representing \textit{decontextualization} that failed to account for the student’s task. Suggestion F provided a logically incorrect interpretation of potential analytic results, representing an \textit{invalid inference}. Suggestion H recommended a rhetorical structure inappropriate for the proposal genre by placing the central hypothesis after extensive background information, representing a \textit{genre mismatch}. In contrast, Suggestions A, B, E, and G provided corresponding sound suggestions concerning theoretical grounding, data collection, analytic approach, and writing clarity and organization, respectively.
The first author manually reviewed a random 10\% sample of the generated feedback (19 responses; 152 AI suggestions) to assess fidelity to the intended experimental design. All flawed suggestions conformed to their intended flaw types. Among sound suggestions, three Suggestion A instances somewhat overstated the theoretical claim while remaining relevant, one Suggestion E instance proposed an incompatible analytical approach, and one Suggestion G instance identified a writing issue that was not present in the draft. All Suggestion B instances conformed to the intended design. Overall, 147 of 152 suggestions (96.7\%) conformed to the intended design, indicating high fidelity of the AI-generated suggestions.

%Following the common workflow of commercial LLM providers, students were then shown all eight suggestions simultaneously and asked to indicate which suggestions they wanted the LLM to implement. The interface included a note stating, ''\textit{AI generated content can include mistakes},'' consistent with common AI writing interfaces that caution users about the possibility of inaccurate output. After students selected the suggestions they wished to adopt, the LLM reprocessed their original draft and returned a revised version incorporating the selected suggestions. Students then individually reflected on the quality of the revisions and the extent to which the revised writing represented their own intended expression. The activity concluded with peer discussion about their experiences using and evaluating AI writing feedback.

Students were then shown all eight suggestions simultaneously and asked to indicate which suggestions they wanted the LLM to implement. This batch presentation follows the workflow of commercial LLM providers, which return revision feedback as an itemized list with a single acceptance question rather than surfacing one issue at a time. We retained this design pattern deliberately to reproduce the practical condition. 
%under which uncritical uptake is most likely in practice. Our comparison condition is more demanding than the commercial default (students had to register a separate accept or reject decision for each of the eight suggestions, rather than applying all of them at once), so the effect of written justification is estimated against a baseline that already imposes some disaggregation. 
Following the AI suggestions, the interface included a note stating, ``\textit{AI generated content can include mistakes},'' consistent with common AI writing interfaces that caution users about the possibility of inaccurate output. After students selected the suggestions they wished to adopt, the LLM reprocessed their original draft and returned a revised version incorporating the selected suggestions. Students then individually reflected on the quality of the revisions and the extent to which the revised writing represented their own intended expression. The activity concluded with peer discussion about their experiences using and evaluating AI writing feedback.

\subsubsection{Randomized Experimental Manipulation}
The randomized experiment was embedded specifically within the decision stage, in which students selected which AI suggestions to adopt. Students were randomly assigned to one of three conditions with approximately equal allocation. The primary experimental comparison involved the \textit{justification} condition and \textit{selection-only} condition. In the justification condition ($n = 65$), students were required to make an accept or reject decision for each suggestion and provide a brief written justification for each decision. In the selection-only condition ($n = 65$), students made a binary accept or reject decision for each of the eight suggestions without the justification requirement. Compared with the default interaction design of most commercial AI writing tools that offers batch acceptance, these conditions required students to make a separate decision for each suggestion, thereby introducing disaggregation into the adoption process. We included a third condition (\textit{batch} condition) as a secondary baseline that approximated the batch interactions ($n = 66$). In this condition, students were first offered options to accept all suggestions, reject all suggestions, or accept only some suggestions. Students who selected ``some'' were then directed to an interface for making individual selections. Due to a technical configuration error, students who selected ``some'' could not select multiple suggestions in the subsequent step. This issue prevented affected students from making informed individual adoption decisions while their initial choices among accepting all, rejecting all, or accepting some remained valid. We therefore used this condition only as a secondary descriptive baseline for examining students' overall selectivity and did not analyze which individual AI suggestions students selected.

Attrition was minimal, with one student assigned to the justification condition and four assigned to the batch condition not completing the experimental task. These responses were removed before analysis. The final analysis included 129 students in the primary comparison between the justification condition ($n = 64$) and the selection-only condition ($n = 65$). An additional 62 students from the batch condition contributed to a supplementary comparison of overall suggestion acceptance selectivity.

\subsubsection{Research Context and Ethics}
The activity was administered through the university’s Qualtrics platform in anonymous mode, and no identifying information was ever collected. The AI feedback and selection activity was first conducted as part of the course’s regular in-class instruction, providing students with hands-on experience that served as a discussion probe for a subsequent class discussion on the use of AI feedback in academic writing. The randomized experiment was embedded within this activity, both for instructors to learn the potential effects of written justification on students’ evaluation of AI feedback and to provide a practical illustration of experimental design before the following week’s lecture on causal inference. After the course concluded, the anonymous activity data were repurposed for research with approval from the university Institutional Review Board.

\begin{table}[t]
\centering
\caption{Design of the eight LLM-generated writing suggestions. Suggestions were deliberately designed to vary in quality, with four sound and four flawed suggestions addressing corresponding aspects of students' proposals. Letters (A-H) indicate the presentation sequence, in which sound and flawed suggestions were interleaved. \textbf{Bolded terms denote LLM flaw types.}}
\label{tab:feedback_design}
\small
\begin{tabularx}{\columnwidth}{@{}l l l X@{}}
\toprule
\textbf{Sugg.} & \textbf{Quality} & \textbf{Feedback focus} & \textbf{Description} \\
\midrule
A & Sound & Theoretical grounding & Suggesting substantively accurate and relevant theoretical support to strengthen or enrich the hypothesis \\
B & Sound & Data source & Suggesting substantively relevant details the student could include about their data collection process \\
C & Flawed & Theoretical grounding & \textbf{Fabrication}: Suggesting fabricated or misapplied theory with fictitious references \\
D & Flawed & Data source & \textbf{Decontextualization}: Suggesting an irrelevant data source to strengthen the analysis \\
E & Sound & Analytic approach & Suggesting relevant and feasible revisions to the analytical approach for testing the hypothesis\\
F & Flawed & Analytic approach & \textbf{Invalid inference}: Suggesting a logically incorrect interpretation of analytic results \\
G & Sound & Writing criteria & Suggesting a revision that addresses a legitimate issue in writing clarity, organization, or grammar\\
H & Flawed & Writing criteria & \textbf{Genre mismatch}: Suggesting an inappropriate proposal structure \\
\bottomrule
\end{tabularx}
\end{table}

\subsection{Analytic Approach}
RQ1 examined how the justification intervention influenced students' selectivity and critical discrimination among AI suggestions. First, we used chi-square tests to examine the extent of non-selective engagement to compare the prevalence of non-selective acceptance, defined as accepting all eight suggestions, across experimental conditions. We also compared decision making time between non-selective adopters and other students within each condition to examine whether non-selective adoption was associated with reduced engagement with the evaluation task. 
Then, we examined whether students' selectivity was sensitive to feedback quality by fitting a mixed effects logistic regression model predicting suggestion acceptance from condition, feedback quality (sound vs. flawed), and their interaction, with student and suggestion identifiers included as random effects to account for the repeated measures structure. We also compared and visualized acceptance rates across conditions for each suggestion. This follow-up analysis helped characterize the pattern underlying the interaction and identify which suggestions contributed to differences in acceptance across conditions.
%Last, to further understand whether students recognized potential problems associated with flawed AI feedback after making their adoption decisions, we conducted follow up regression analyses examining whether the numbers of accepted helpful and unhelpful suggestions were associated with students' perceptions of their drafts following the activity. We constructed two linear regression models predicting students' perceived authorship and perceived writing quality, respectively, from the numbers of accepted helpful and unhelpful suggestions, while controlling for self-reported AI use and experimental condition.

RQ2 focused on gaps in students' critical evaluation of flawed AI suggestions as reflected in their written justifications. Although critical evaluation can involve both identifying flaws in problematic suggestions and recognizing the value of helpful ones, our analysis focused on the former because the study design deliberately embedded identifiable flaws in selected AI suggestions. This established a basis for examining whether and how students recognized these flaws in their decisions and justifications. In contrast, whether a sound suggestion is helpful may depend on students' individual writing goals and preferences, making its appropriateness context dependent. We therefore focused RQ2 on students' evaluations of flawed AI suggestions, for which their decisions and justifications could be interpreted against established flaws in the suggestions.
To address RQ2, reflexive thematic analysis was conducted following Braun and Clarke's framework ~\citep{braun2006using, braun2019reflecting} to develop themes from students' written justifications. Unlike coding reliability approaches, which emphasize consistent coding across researchers and treat coder consensus as evidence of analytic validity, reflexive thematic analysis views themes as interpretive constructions generated through researchers' active engagement with and interpretation of the data~\citep{byrne2022worked}. Multiple researchers are therefore not required but may contribute to reflexive thematic analysis by enriching and challenging interpretations rather than establishing consensus as evidence of a single correct interpretation. In our analysis, the first author took the primary role in conducting the reflexive thematic analysis. The analysis began with the first author reviewing all written justifications to develop familiarity with the dataset and its range of responses. The first author then wrote a brief analytic note summarizing the main point of each justification, which served as the basis for initial coding. The first author subsequently reviewed each student's original proposal draft and the corresponding LLM suggestion to contextualize and refine the interpretation of students' justifications, particularly when responses were brief or ambiguous. Initial codes were then compared and iteratively grouped into themes representing recurring patterns in students' demonstrated reasoning. Themes were examined in relation to one another and refined to clarify their conceptual boundaries. For example, the initially distinct codes \textit{Scope Fit}, referring to the rejection of a suggestion because it fell outside the project's intended scope, and \textit{Topic Irrelevance}, referring to rejection because the suggestion was unrelated to the proposal's topic, were merged into a broader theme. The resulting themes, names, and definitions were then reviewed and discussed among the authors. Through these discussions, the other authors questioned and contributed to the refinement of the first author's interpretations, and the thematic structure was subsequently finalized.

\subsubsection{Positionality and Reflexivity Statement}

As an educational researcher with extensive teaching experience in college-level data science courses, the first author approached the thematic analysis with prior familiarity with data science knowledge and research on students' overreliance on AI. This background provided useful domain expertise for interpreting students' justifications. Epistemologically, the analysis adopted an experiential orientation, treating students' written justifications as accounts that provide insight into their evaluative reasoning while recognizing that such accounts are partial and context dependent. Consistent with this orientation, the analysis was primarily inductive: rather than applying a predetermined taxonomy of reasoning, the first author developed codes and themes through engagement with recurring patterns in students' justifications. This approach allowed interpretations to remain grounded in students' own accounts while still reflecting the researchers' active role in constructing meaning from the data.

Throughout the analysis, the first author reflected on how prior knowledge of flaws in the AI suggestions could lead to overinterpreting students' justifications, particularly by attributing recognition of a flaw when it was not explicitly expressed. To mitigate this risk, the first author iteratively revisited the data with varying levels of contextual information. Initial coding focused solely on students' justifications, followed by contextual review of the corresponding proposal and AI suggestion to refine interpretations of ambiguous responses. The first author then returned to the justifications, comparing interpretations across rounds and across similarly worded responses to identify instances where knowledge of the AI suggestion may have led to inferences not sufficiently supported by students' responses. 

\section{Results}
\subsection{Overview of Student Proposal Drafts}

Before the experiment, students completed an initial draft of a proposal essay, which served as the starting point for the experimental activity. Among the responses included in the study, these first proposal drafts averaged 358.9 words in length, ranging from 81 to 1,255 words. Students spent an average of 12.28 minutes drafting their proposals. Seventy responses (54.3\%) indicated that students did not use AI during the drafting process, based on their self-reports, whereas 59 responses (45.7\%) indicated AI use. Among the responses indicating AI use, 39 involved using AI to improve the writing, 9 to conduct research or generate ideas, 6 to generate part of the essay, 2 to generate the entire essay, and 3 for other purposes. There were no significant differences across experimental conditions in draft length, time spent drafting, or self-reported AI use, indicating that the conditions were comparable on these pre-experimental characteristics.

\subsection{RQ1: Justification Increases Discrimination Between Sound and Flawed Suggestions}
During the experiment, each participant received 8 mixed-quality revision suggestions from an LLM. We examined the frequency of non-selective acceptance, defined as accepting all eight suggestions. In the justification condition, $10$ responses ($15.6\%$) reflected non-selective acceptance. These responses spent significantly less time making their adoption decisions than other responses in the same condition (median = 122.6 vs. 296.6 seconds, Mann-Whitney $U = 529$, $p = .003$). In the selection-only condition, $20$ responses ($30.8\%$) reflected non-selective acceptance and also spent significantly less time making their adoption decisions than other responses in the same condition (median = 53.3 vs. 120.5 seconds, $U = 743$, $p < .001$). As a secondary descriptive baseline, 20 of 62 students ($32.3\%$) in the batch condition selected the option to accept all suggestions. This provides contextual evidence that the rate of wholesale acceptance observed in the selection-only condition was similar to that observed when students were given a less disaggregated, batch style decision structure. Non-selective acceptance was less frequent in the justification condition, with evidence of a significant difference compared to the batch condition ($\chi^2(1) = 3.842$, $p = .050$) and a similar directional pattern compared to the selection-only condition ($\chi^2(1) = 3.339$, $p = .068$). 

We further examined whether the increased selectivity in the justification condition effectively discriminated between sound and flawed AI suggestions. As shown in Table~\ref{tab:regression}, the generalized linear mixed effects model random intercepts for students and suggestion items revealed a significant interaction between experimental condition and suggestion quality ($\beta = -1.713$, $SE = 0.351$, $OR = 0.180$, $p < .001$), indicating that the effect of requiring written justification differed substantially by suggestion quality. We also fitted a model with a by-student random slope for suggestion quality. The significance of the $condition \times quality$ interaction was unchanged ($\beta = -1.982$, $SE = 0.496$, $OR = 0.138$, $p < .001$), and we report the simpler model here. The interaction OR of 0.180 indicates that the odds ratio for the effect of written justification was 0.18 times as large for flawed suggestions as for sound suggestions. Specifically, for sound suggestions, the justification intervention was associated with slightly higher odds of acceptance, although the difference was not statistically significant ($OR = 1.491$, $p = 0.265$). In contrast, for flawed suggestions, the estimated odds of acceptance under written justification were 0.269 times those in the selection-only condition, corresponding to approximately 73\% lower odds of acceptance.

\begin{figure*}[t]
    \centering
    \includegraphics[width=0.9\textwidth]{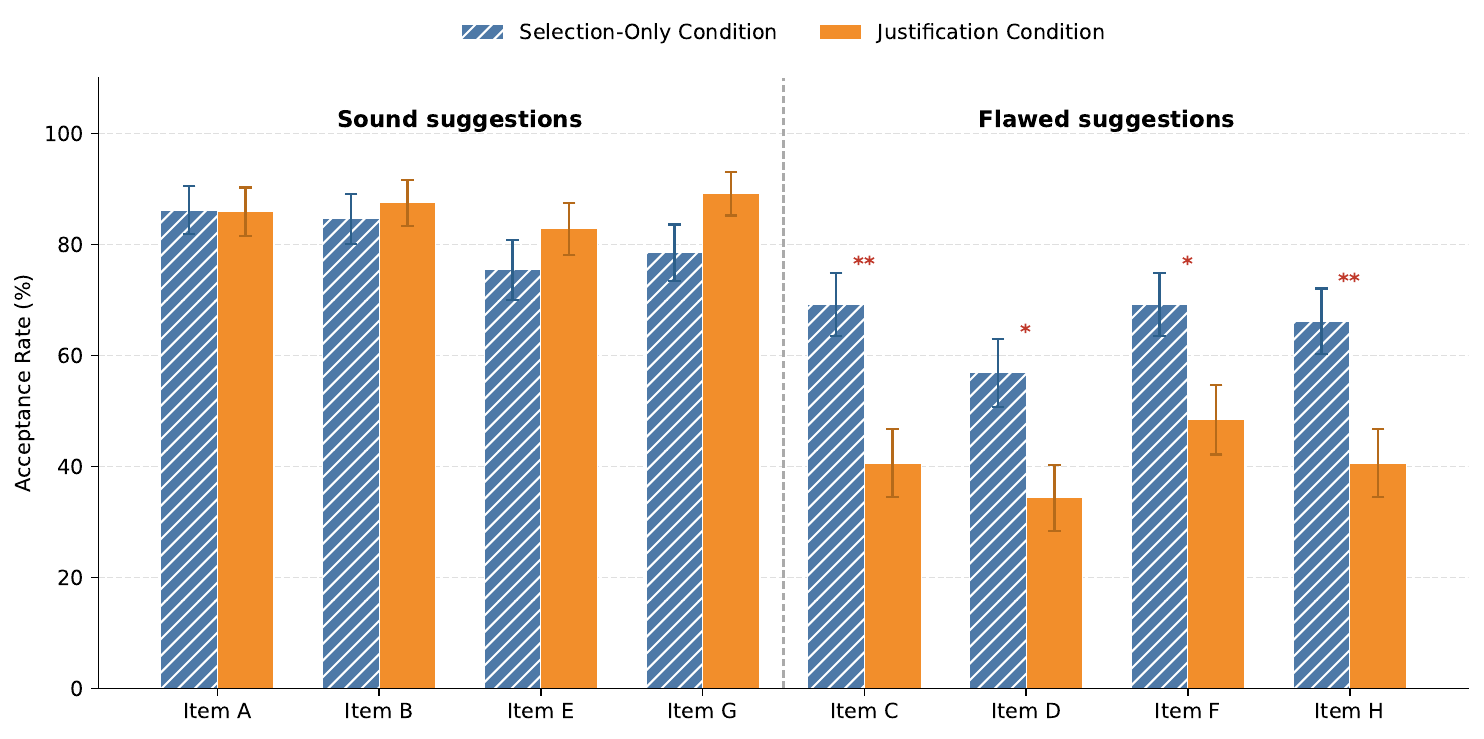}
    \caption{Acceptance Rates (\%) for Sound and Flawed AI-Generated Writing Suggestions by Condition. Blue bars with diagonal lines represent the Selection-Only condition ($n = 65$) and solid orange bars represent the Justification condition ($n = 64$). Asterisks indicate statistically significant differences in acceptance rates between conditions based on chi-square tests (* $p < .05$, ** $p < .01$, *** $p < .001$); items without asterisks showed no significant difference between conditions.}
    %\Description{A grouped bar chart showing acceptance rates for eight AI-generated writing suggestions (Items A, B, C, D, E, F, G, H) across two experimental conditions. The x-axis lists the eight items, divided by a dashed vertical line into sound suggestions (Items A, B, E, G, on the left) and flawed suggestions (Items C, D, F, H, on the right). The y-axis shows acceptance rate as a percentage from 0 to 100. For each item, two bars are shown side by side: a blue hatched bar for the Selection-Only condition (n = 65) and a solid orange bar for the Justification condition (n = 64). For sound suggestions, acceptance rates are high and similar across both conditions, ranging from approximately 75\% to 89\%, with no statistically significant differences between conditions. For flawed suggestions, the justification condition shows markedly lower acceptance rates than the Selection-Only condition: Item C (40.6\% vs. 69.2\%), Item D (34.4\% vs. 56.9\%), Item F (48.4\% vs. 69.2\%), and Item H (40.6\% vs. 66.2\%). All four flawed items are marked with asterisks indicating statistically significant between-condition differences (** $p < .01$ for Items C and H; * $p < .05$ for Items D and F).
%}
    \label{fig:acceptance_rate}
\end{figure*}

Figure~\ref{fig:acceptance_rate} presents acceptance rates separately by suggestion quality to further characterize the $condition \times quality$ interaction. As follow-up analyses, we examined condition differences separately for each suggestion. For all four flawed suggestions, acceptance was lower in the justification condition: Suggestion C ($\chi^2 = 9.54$, $p = .002$), Suggestion D ($\chi^2 = 5.73$, $p = .017$), Suggestion F ($\chi^2 = 4.93$, $p = .026$), and Suggestion H ($\chi^2 = 7.45$, $p = .006$). Acceptance of sound suggestions did not significantly differ between conditions. The average acceptance rate of flawed suggestions decreased from 65\% to 41\% with required decision justification, with no corresponding reduction in the average acceptance of sound suggestions (81\% vs. 86\%). These results suggest that requiring written justification increased students' selectivity primarily by reducing their acceptance of flawed AI suggestions while preserving their acceptance of sound suggestions.

\begin{table}[t]
\centering
\caption{Generalized linear mixed effects model predicting students' acceptance of AI-generated writing suggestions.}
\label{tab:regression}
\begin{threeparttable}
\small
\begin{tabular}{lcccc}
\toprule
 & $\beta$ & 95\% CI ($\beta$) & OR & 95\% CI (OR) \\
\midrule
\multicolumn{5}{l}{\textbf{Fixed Effects}} \\
Intercept
    & 1.927$^{***}$
    & [1.409, 2.444]
    & 6.866
    & [4.091, 11.523] \\
    & (0.264) & & & \\[2pt]

Condition (Justification)
    & 0.399
    & [$-0.303$, 1.102]
    & 1.491
    & [0.738, 3.009] \\
    & (0.358) & & & \\[2pt]

Quality (Flawed)
    & $-1.064^{***}$
    & [$-1.560$, $-0.568$]
    & 0.345
    & [0.210, 0.567] \\
    & (0.253) & & & \\[2pt]

Condition $\times$ Quality
    & $-1.713^{***}$
    & [$-2.400$, $-1.026$]
    & 0.180
    & [0.091, 0.358] \\
    & (0.351) & & & \\

\midrule
\multicolumn{5}{l}{\textbf{Random Effects}} \\
Student intercept variance
    & 1.653 & & & \\
Item intercept variance
    & 0.016 & & & \\

\midrule
\multicolumn{5}{l}{\textbf{Model Information}} \\
Observations
    & 1,032 & & & \\
Students
    & 129 & & & \\
Items
    & 8 & & & \\
AIC
    & 1,070.8 & & & \\
BIC
    & 1,100.4 & & & \\
Log likelihood
    & $-529.4$ & & & \\
\bottomrule
\end{tabular}

\begin{tablenotes}
\footnotesize
\item \textit{Note.} $\beta$ represents log odds coefficients. OR represents odds ratio, while the OR for the interaction term represents a ratio of odds ratios. 
Standard errors are reported in parentheses below the corresponding
$\beta$ estimates. 95\% CIs are Wald confidence intervals. The reference
groups are the selection-only condition and sound AI suggestions. Random
intercepts were included for students and suggestion items. $^{*}p<.05$,
$^{**}p<.01$, $^{***}p<.001$.
\end{tablenotes}
\end{threeparttable}
\end{table}

\subsection{RQ2: Justification Reveals Persistent Gaps in Effectively Evaluating AI Suggestions}
%The written justification recorded in the experimental condition also revealed persistent gaps in students' critical evaluation of AI wiring suggestions. 
Table~\ref{tab:themes} summarizes the themes generated from students' written justifications for accepting or rejecting flawed LLM suggestions in the justification condition.

Among justifications for accepting flawed AI suggestions, vague endorsement without a specific rationale was the most common theme for all flawed suggestions. These justifications were typically very brief, endorsing the perceived benefits of the AI suggestion without explaining why it was appropriate for the student's proposal. For example, students wrote: ``\textit{stronger evidence}'', ``\textit{worth adding}'', ``\textit{it's reasonable.}''

Heuristic reasoning was another common theme in students' acceptance of flawed suggestions. These justifications recognized a general principle underlying the AI suggestion and endorsed it based on a broad heuristic that a particular type of addition or revision is generally beneficial, without evaluating the specific suggestion in the context of their own proposal. For example, for Suggestion C, which provided fabricated theoretical support, students wrote, ``\textit{extra evidence always nice}'', ``\textit{objectively it would be best to have existing studies referenced in my hypothesis.}'' For Suggestion D, which recommended a decontextualized data source, students assumed that expanding the data source or scope would automatically increase statistical power, rigor, or credibility, as illustrated by the justification: ``\textit{Yes using an extra data source could help support the findings and make the results look more credible.}''. For Suggestion F, which proposed a logically incorrect interpretation of potential results, students similarly endorsed the general value of interpreting or examining statistical results. They wrote, ``\textit{This is good because possible results should be incorporated in the proposal}'', and ``\textit{Yes, statistical differences should be looked at.}'' 

Acceptance justifications also sometimes echoed the AI suggestion's rhetorical rationale rather than independently evaluating its appropriateness. This pattern was particularly evident for Suggestion H, which recommended a genre-mismatched writing structure. Students repeated the suggestion's promised benefits, such as improving flow, readability, or structure, without examining whether the proposed structure was appropriate for their proposal. Only a small number of justifications for flawed suggestion acceptance engaged more deeply with the relationship between the suggestion and their own arguments, although these students still failed to identify the specific flaw in the AI suggestion.

Among justifications for rejecting flawed AI suggestions, the extent to which students correctly identified the underlying AI flaw varied across flaw types. For the flaw of invalid inference (Suggestion F), rejection justifications were dominated by correct detection of the logical flaw in AI reasoning. In contrast, for the more context dependent flaws of decontextualization (D) and genre mismatch (H), rejection justifications frequently reflected two themes: correctly identifying the contextual problem, and judging that the suggested change was ``\textit{unnecessary}''. The ``\textit{unnecessary}'' theme is particularly noteworthy because it encompass qualitatively different forms of reasoning. Some justifications explicitly cited what the student believe they had already achieved, the perceived genre convention, task requirements, or writing goals. For example, students wrote, ``\textit{For D, I think it’s optional. Including a national dataset could make the study more generalizable, but for a short proposal, it might be unnecessary. I might mention it briefly as future work, but I wouldn’t focus on it}.'' In contrast, other justifications invoking ``unnecessary'' were brief and unelaborated (e.g., ``\textit{I don't need it}'', ``\textit{Not really necessary}'', ``\textit{I don’t think this is that necessary}''). These unelaborated justifications could reflect similarly careful evaluations that students did not articulate, but they could also represent general or intuitive judgments that the suggested change was not needed. Besides, some of them may further reflect limited engagement with the justification requirement, as some students used exactly the same justification ``\textit{Not really necessary}'' across all flawed suggestions. 

For the flaw of fabrication (Suggestion C), students' justifications revealed limitations in their attention to the factual validity of AI generated claims. The most common reason cited in students' written justification was the suggested theory contradicts or is irrelevant to the proposal hypothesis. Among those responses ($n = 24$), 18 gave this as the sole reason for rejection, without recognizing that the theory itself was fabricated or misrepresented. For example, students wrote, ``\textit{I feel like if I'm arguing for the attempt-first side, why would I propose evidence that goes against that?}'' This pattern suggests that students could focus on argumentative evaluation of whether a suggested theory fit their own argument while overlooking the more fundamental question of whether the suggested theory was factually valid. 

Only 14 of the responses rejecting suggestion C explicitly mentioned the factual error. Among those responses, we observed differences in how students cast doubts on the truthfulness of the AI suggestion. Three patterns were evident. First, some responses explicitly indicated direct fact or source auditing, showing the student actively attempted to verify the existence or validity of cited theories, sources, or journal names. For example, one student wrote, ``\textit{This theory doesn't appear to be real. [The suggested theory] cannot be verified and the journal name itself sounds fabricated.}'' Second, some students relied on domain knowledge sanity check, drawing on prior subject matter knowledge to identify claims that appeared inconsistent with established knowledge. For example, students wrote, ``\textit{This feedback is incorrect, it contradicts established learning theory}'', and ``\textit{No, because the theory mentioned is not well established and contradicts stronger evidence that active effort improves learning.}'' Third, some students expressed \textit{skepticism}, recognizing uncertainty about the AI's claim and withholding acceptance pending further verification. For example, one student wrote, ``\textit{not really sure if it is true, require further reading and investigation}.'' These patterns illustrate that students' fact checking for AI suggestions varied not only in whether they detected potential factual problems, but also in the strategies they used to evaluate the truthfulness of AI outputs.

\begin{table}[ht]
\centering
\caption{Themes in Students' Justifications for Accepting or Rejecting Flawed AI Suggestions. Themes are not mutually exclusive; a justification could be assigned to more than one theme, so totals exceed the number of responses. The analysis covers 256 justifications for flawed suggestions from 64 students.}
\label{tab:themes}
\begin{tabular}{@{}p{0.16\columnwidth} p{0.07\columnwidth} p{0.60\columnwidth} p{0.04\columnwidth}@{}}
\toprule
\textbf{Flaw} & \textbf{Decision} & \textbf{Theme} & \textbf{N}\\
\midrule

\multirow{9}{*}{Fabrication} 
    & \multirow{4}{*}{Accepted} 
    & Vague endorsement without a specific rationale & 12 \\
    & & Heuristic reasoning that additional theoretical sources or citations strengthen the argument & 9 \\
    & & Belief that the suggested theory fits or contributes to the argument & 5 \\
\cmidrule{2-4}
    & \multirow{5}{*}{Rejected} 
    & Recognition that the suggested theory contradicts or is irrelevant to the argument & 24 \\
    & & Detection of factual errors in the suggested theory or references & 14 \\
    & & Judgment that the suggested theory is unnecessary & 4 \\
    & & Rejection or disagreement without a specific rationale & 3 \\
\midrule

\multirow{10}{*}{Decontextualization} 
    & \multirow{4}{*}{Accepted} 
    & Vague endorsement without a specific rationale & 10 \\
    & & Belief that expanding the data source or scope would strengthen the analysis & 10 \\
    & & Belief that the suggested data source fits or contributes to the proposal & 2 \\
\cmidrule{2-4}
    & \multirow{6}{*}{Rejected} 
    & Recognition that the suggested data source falls outside the topic or scope of the proposal & 20 \\
    & & Judgment that the suggested data source is unnecessary & 16 \\
    & & Recognition that the suggested data source is incompatible with the proposed analysis & 4 \\
    & & Judgment that the suggested data source is impractical or infeasible & 4 \\
\midrule

\multirow{11}{*}{Invalid inference} 
    & \multirow{6}{*}{Accepted} 
    & Vague endorsement without a specific rationale & 15 \\
    & & Perception that the suggested interpretation would improve rhetorical clarity or specificity & 7 \\
    & & Appreciation for the methodological details or information provided & 6 \\
    & & Heuristic reasoning that interpreting or examining statistical results strengthens the proposal & 3 \\
\cmidrule{2-4}
    & \multirow{5}{*}{Rejected} 
    & Detection of logical flaws in the suggested interpretation & 22 \\
    & & Judgment that adding the suggested interpretation is unnecessary & 8 \\
    & & Perception that the suggested interpretation does not fit or is irrelevant to the argument & 2 \\
    & & Rejection or disagreement without a specific rationale & 2 \\
\midrule

\multirow{12}{*}{Genre mismatch} 
    & \multirow{5}{*}{Accepted} 
    & Vague endorsement without a specific rationale & 13 \\
    & & Echoing the suggestion's rhetorical rationale about improving flow, readability, or structure & 7 \\
    & & Heuristic preference for presenting contextual information before the hypothesis & 6 \\
\cmidrule{2-4}
    & \multirow{7}{*}{Rejected} 
    & Judgment that the suggested structural change is unnecessary & 12 \\
    & & Preference for a different proposal structure based on genre conventions & 9 \\
    & & Concern that the suggested structure would weaken the paper's rhetorical effectiveness & 9 \\
    & & Recognition that the suggested structure does not fit the task requirements or constraints & 8 \\
    & & Rejection or disagreement without a specific rationale & 5 \\
\bottomrule
\end{tabular}
\end{table}

\section{Discussion}
Uncritical uptake of AI suggestions was common in our study, not only where the interface invited this behavior. Nearly one third of students in the batch condition, which approximated the interaction design of commercial AI writing tools, accepted all AI suggestions wholesale. Almost as many did so when students were required to make a separate decision for each suggestion (32\% vs. 31\%). This echoes students' self-reported experiences of uncritical AI uptake in prior survey and interview research~\citep{du2026exploring, mujeeb2026beyond}, and adds behavioral evidence that students may still accept AI suggestions wholesale even when they are reminded that AI can make mistakes.

Going into the study, we expected that requiring suggestions to be judged individually would improve critical uptake. If the wholesale acceptance model promoted in commercial AI tools were mainly a product of the batch acceptance interface, then requiring a separate decision for each suggestion should have reduced it. Our findings, however, suggest that it did not. Disaggregating judgments in the interface changes what the interaction costs, but it does not by itself change how students decide. Being asked for reason is what actually improved critical uptake. Thus, the effective ingredient is demanding an articulation of reasoning, not merely increasing the number of clicks. This distinction matters for interface design: adding steps to an interface is not the same as adding deliberation to a decision when the goal is to promote critical thinking.

Our findings also showed the reduced non-selective acceptance with required justification for suggestion adoption decisions did not reflect a general reduction in AI acceptance, but greater discrimination between sound and flawed suggestions, with students becoming less likely to accept suggestions containing flaws. These findings provide causal evidence that written justification can serve as an effective, lightweight intervention for mitigating students' uncritical uptake of AI suggestions during writing revision. A key advantage of this intervention is that it can be embedded directly into students' interactions with AI. Rather than replacing broader instruction or training in critical thinking and AI literacy \citep[e.g.,][]{hou2026effects}, written justification can complement these approaches by prompting students to exercise critical evaluation during their interactions with a specific AI suggestion. Broader interventions may help develop the knowledge and skills needed for critical AI use, while embedded justification provides a lightweight mechanism for activating and applying those capacities during actual human-AI interaction. When incorporated throughout the learning process, such repeated opportunities for evaluation may also help habituate critical evaluation as a routine practice in students' everyday interactions with AI~\citep{vendrell2026scaffolding}.

It is worth noting that the justification intervention fell short of eliminating uncritical uptake of AI suggestions in academic writing. Around 15\% of students in the justification condition still accepted all eight suggestions, and flawed suggestions were still 
%demonstrated non-selective adoption of mixed-quality AI suggestions. Increased selectivity in suggestion adoption also did not completely eliminate problematic acceptance of flawed suggestions. With the justification intervention, Increased selectivity also did not eliminate problematic acceptance of flawed suggestions, which continued to be 
accepted at an average rate of 41\%. These findings highlight both the limitations of the current intervention design and opportunities to strengthen it. Our analysis of students' written justifications provides insights into these opportunities. 

First, substantive engagement with the justification task emerged as an important challenge. Responses that accepted all suggestions were typically accompanied by short, superficial justifications, sometimes including the same brief justification repeated across multiple suggestions. Similarly, vague endorsement without elaborated reasoning was one of the most common patterns among justifications for accepting flawed suggestions. Among justifications for rejecting flawed suggestions, a substantial category simply described a suggestion as ``unnecessary.'' While such responses may reflect careful evaluation of the suggestion's relevance or value to the student's proposal, they may also reflect limited engagement with the evaluation and justification process, particularly when the response provided no further explanation. Together, these findings suggest that requiring justification alone does not guarantee meaningful engagement with evaluation. When implementing this intervention in instructional activities or AI writing tools, instructors and technology designers should consider how to encourage substantive engagement rather than allowing justification to become merely a procedural requirement. For example, designers might increase the perceived value of the justification task by explaining its role in learning or incorporating students' reasoning into assessment and feedback~\citep{wigfield2000expectancy}. At the same time, implementation should account for the perceived cognitive and time demand of the task and avoid overwhelming students with excessive justification requirements~\citep{yeh2010optimal}. Designing an effective intervention may therefore require balancing sufficient cognitive engagement with a level of effort that students can realistically sustain.

Students' justifications also revealed gaps in students' metacognitive monitoring of AI suggestions. Students sometimes relied on heuristic reasoning or simply echoed the rhetorical promises of an AI suggestion without carefully examining the truthfulness of its content or considering how the proposed change would affect their existing draft. There was also an evident gap in students' monitoring of the factual truthfulness of AI suggestions. These findings indicate that while the justification task adds friction to the decision making process, it may not provide sufficient guidance to activate and direct the metacognitive processes needed for effective evaluation. Prior research suggests that metacognitive prompts containing specific monitoring and reflection questions can support more deliberate and critical use of AI tools~\citep{singh2025enhancing}. Future implementations should therefore combine written justification with more explicit metacognitive scaffolding, such as prompts asking students what aspects of a suggestion they should evaluate, what criteria they should use, and what evidence they should seek before accepting it. 

The written justifications also revealed gaps in students' domain knowledge that may impede effective evaluation of AI suggestions. For example, some students rejected Suggestion C because it introduced a counterargument, reflecting a potentially concerning misconception that counterarguments should be avoided in academic writing, even though engaging with alternative perspectives is an important rhetorical practice in academic argumentation. Although these students correctly rejected the flawed AI suggestion, their justifications revealed reasoning that could lead them to reject appropriate suggestions in other contexts. Conversely, the responses of students who correctly identified the fabrication in Suggestion C drew on their prior knowledge of learning theories, further illustrating the role of domain knowledge in evaluating AI suggestions. Together, these findings highlight the role of existing domain knowledge in shaping students' evaluation of AI suggestions, raising a critical question about who benefits from interventions that introduce friction. Justification gives students an opportunity to apply what they know, but does not necessarily support them around what they do not know. If the students best positioned to catch a fabricated citation are those who already know the relevant literature, then friction of this kind will mostly help the already-prepared. Such interventions could therefore widen rather than narrow gaps in how well students use AI feedback. Testing this possibility directly would require measuring students' domain knowledge before the activity, which was not possible in our anonymous study design. Future research should examine this question and investigate how instructional support during and beyond AI interactions can better support students who lack the domain knowledge needed to critically evaluate AI suggestions.
%Thus, to facilitate effective AI supported writing, instructors may need to combine justification based evaluation with instruction that develops students' domain knowledge and understanding of effective writing practices.

%This study have several limitations and points to directions for future research. First, the experiment was conducted within a single course and focused on one academic writing context. Future studies should examine whether the findings generalize to students working on other types of writing, in other disciplines, and in other instructional settings. Second, the intervention targeted a specific point in the writing process: students' decisions about whether to accept or reject AI generated revision suggestions. We did not examine how students formulate prompts for obtaining AI feedback, how they respond to or ``talk back'' to AI suggestions, or how they develop an understanding of what makes writing effective in the first place. These stages may involve different forms of metacognitive monitoring and control and may require different forms of instructional support. Last, because the present activity was administered anonymously, we could not definitively determine whether each completed response corresponded to a unique student or rule out repeated participation. Future research could use confidential participant identifiers or linked pre- and post-measures to verify unique participation while preserving students' privacy.

\subsection{Limitations and Future Work}

Our study has several limitations that bound the implications of our findings. First, the design cannot separate the effect of justification from the effect of the time and effort it requires to produce them. Writing a reason takes longer than clicking a checkbox and being asked for a reason may itself signal that some suggestions are worth doubting. Either mechanism could produce the observed pattern. However, two features of our results argue against a pure suspicion-cueing explanation: the intervention did not lower acceptance of sound suggestions and students discriminated between suggestion types rather than rejecting more overall. A cleaner (but ecologically less valid) test would hold time on task constant by adding a condition that requires an equivalent but non-evaluative written response.

Second, the suggestion set was more adversarial than students would encounter in real-world use of an AI tool for the purpose of writing feedback. Half of the eight suggestions had a deliberate flaw, which is well above the rate at which current models produce them. We chose this ratio to make gaps in evaluation observable within a single session with sufficient statistical power. As a result, the absolute acceptance rates in our study should not be read as estimates of real-world uptake. Only the comparison between conditions is meaningful given our design.

Third, adoption decisions in this study are a proximal outcome. We did not assess whether the revised drafts were better. A student who rejects a flawed suggestion for the wrong reason (which we observed frequently) counts the same in our data as one who diagnoses the flaw. The activity was also ungraded, which lowers the incentive for students to invest effort in a justification and may understate the effect of adding a justification requirement under graded assessment conditions.

Fourth, our findings come from a single course and one academic writing context. Whether they generalize to other disciplines, other genres, and other instructional settings is an open question. The intervention also only targeted one point in the writing process. We did not examine how students formulate prompts, or how they would respond to suggestions (in an conversational AI tool). These steps may involve different forms of metacognitive monitoring and require different instructional support.

Finally, we treated each Qualtrics response as representing a unique student. Because the activity was administered anonymously, however, we could not verify that each response came from a distinct student, although teaching assistants were present to monitor students' participation in the activity. Future work could use confidential participant identifiers or linked pre- and post-measures to establish this while preserving student privacy.

\section{Conclusion}
This study examined the use of strategic friction to mitigate uncritical uptake of AI-generated revision suggestions in academic writing by requiring students to justify their adoption decisions. Through a randomized experiment, we provide causal evidence that this lightweight intervention can increase students' selective adoption of AI suggestions by improving their discrimination between sound and flawed suggestions. At the same time, written justification is no panacea for uncritical uptake or problematic acceptance of flawed AI suggestions. Our findings also revealed challenges in the intervention's effectiveness, including superficial engagement with the justification task and gaps in students' metacognitive monitoring and domain knowledge that can limit effective evaluation of AI suggestions. Together, these findings highlight both the potential of embedding strategic friction directly into human--AI interactions and the need for complementary support to foster deeper and more effective critical engagement with AI.

\bibliography{ref}
\bibliographystyle{iclr2027_conference}

%\appendix
%\section{Appendix}
%You may include other additional sections here.

\end{document}